\documentclass[backend=biber]{gistt}
\usepackage{authblk}
\usepackage{siunitx}
\usepackage{paralist}
\usepackage[pdftex]{graphicx}
  \graphicspath{{./figures/}}
  \DeclareGraphicsExtensions{.pdf,.tex,.png}
\usepackage{tikz}

\definecolor{Paired0}{RGB}{166,206,227}
\definecolor{Paired1}{RGB}{ 31,120,180}
\definecolor{Paired2}{RGB}{178,223,138}
\definecolor{Paired3}{RGB}{ 51,160, 44}
\definecolor{Paired4}{RGB}{251,154,153}
\definecolor{Paired5}{RGB}{227, 26, 28}
\definecolor{Paired6}{RGB}{253,191,111}
\definecolor{Paired7}{RGB}{255,127,  0}
\definecolor{myRed}{RGB}{228,26,28}

\newcommand\myline[1]{%
  \begin{tikzpicture}[baseline, color=myRed]
    \draw[very thick,#1](0,\dp\strutbox-1)--(\ht\strutbox,\dp\strutbox-1);%
  \end{tikzpicture}%
}

\newcommand\fourlines[4]{%
  \begin{tikzpicture}[baseline,solid,very thick]
    \def\varOffsetW{-4}%
    \draw[color=#1](0,\dp\strutbox+2)--(\ht\strutbox+\varOffsetW,\dp\strutbox+2);%
    \draw[color=#2](0,-\dp\strutbox+7)--(\ht\strutbox+\varOffsetW,-\dp\strutbox+7);%
    \draw[color=#3](0,-\dp\strutbox+5)--(\ht\strutbox+\varOffsetW,-\dp\strutbox+5);%
    \draw[color=#4](0,-\dp\strutbox+3)--(\ht\strutbox+\varOffsetW,-\dp\strutbox+3);%
  \end{tikzpicture}%
}

\newcommand\float[2][round-precision = 2]{%
  \num[round-mode = places,
  scientific-notation = fixed, fixed-exponent = 0,
  output-decimal-marker={.}, #1]{#2}%
}

\title{Detection of Performance Changes in MooBench Results \\ Using Nyrki\"{o} on GitHub Actions}

\author[1]{Shinhyung Yang}
\author[2]{David Georg Reichelt}
\author[3]{Henrik Ingo}
\author[1]{Wilhelm Hasselbring}

\affil[1]{Kiel University, Kiel, Germany}
\affil[2]{Lancaster University Leipzig / URZ Leipzig, Leipzig, Germany}
\affil[3]{Nyrki\"{o} Oy, J\"{a}rvenp\"{a}\"{a}, Finland}

\begin{document}

\maketitle

\begin{abstract}

In GitHub with its 518~million hosted projects, performance changes within these
projects are highly relevant to the project's users. Although performance
measurement is supported by GitHub CI/CD, performance change detection is a
challenging topic.

In this paper, we demonstrate how we incorporated Nyrki\"{o} to MooBench.
Prior to this work, Moobench continuously ran on GitHub virtual machines,
measuring overhead of tracing agents, but without change detection. By adding
the upload of the measurements to the Nyrki\"{o} change detection service, we
made it possible to detect performance changes.
We identified one major performance regression and examined the performance change in depth. We report that (1) it is reproducible with GitHub actions, and (2) the performance regression is caused by a Linux Kernel version change.

\end{abstract}

\section{Introduction}

GitOps is a specific implementation and extension of DevOps practices,
particularly focused on using Git as the single source of truth for
infrastructure and application deployments. GitOps provides a set of development
operations that embody the CI/CD tasks~\cite{Beetz2022}. GitOps integrates
DevOps operations of a software project with Git operations, e.g., with GitOps,
a git-push action triggers a user-defined CI/CD pipeline, specific to the
software project's repository.

GitHub has become the biggest GitOps platform for day-to-day development in
open-source and enterprise projects.  GitHub reports that it hosts 518~million
projects, including one billion
contributions.
\emph{GitHub Actions}
is a
CI/CD platform that includes both CI/CD pipelines and \emph{GitHub runners}:
Azure VM resources that run actions.  The github-action-benchmark
tool\footnote{\url{https://github.com/benchmark-action/github-action-benchmark}}
is a continuous benchmarking tool for GitHub CI/CD; it is provided as a
\emph{GitHub action}, which receives the benchmarking data, and plots the result
on a GitHub.io page. It is useful for comparing performance differences between
git pushes, where the performance is not only impacted by software changes, but
also by the VMs.

Benchmarks can be used for comparing different methods, techniques and
tools~\cite{EASE2021}, and MooBench~\cite{Waller2015, Reichelt2021} is used for
comparing tracing agents of monitoring frameworks such as
Kieker~\cite{Kieker2025}.

In this paper, we demonstrate that performance change detection of
MooBench's measurement results works by running the Nyriö change detection service, integrated with MooBench's GitHub actions workflow. Nyrki\"{o} is
a change detection service provided as a GitHub action. It uses the E-Divisive
Means algorithm to detect changes~\cite{EDivisive2020}. In the examined
versions, we see that the execution time of Kieker's AspectJ tracing agent increased
significantly on GitHub's {\tt ubuntu-latest} image, although the
involved commits did not contain any changes to the software. We confirmed the
detected change by reproducing and analyzing the performance change locally.

Our contributions for this paper are as follows:
\begin{inparaenum}[(1)]
  \item we integrated the Nyrki\"{o} change detection service to continouously
  analyse the MooBench benchmarking,
\item we discovered and replicated the performance change on GitHub runners, and
\item we configured GitHub-hosted and self-hosted runners for replicating the results.
\end{inparaenum}

\section{Background}

\begin{figure}[t]
  \centering
  \input{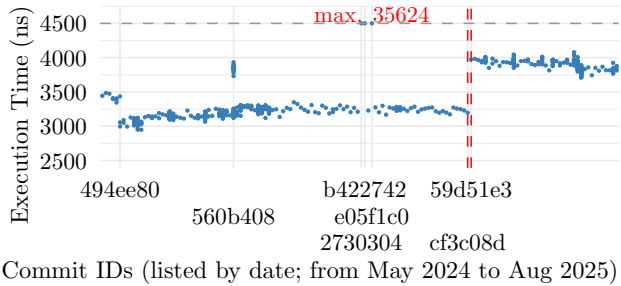}
  \vspace{-1.40em}
  \caption[Commit Diagram]
  {%
    Our investigation focuses on the change between {\tt cf3c08d} and {\tt
    59d51e3} (\myline{dashed}). Each data point represents the execution time of
    a Benchmark run. Values $>$ \SI{4500}{\nano\second} are clipped and the
    maximum value \num{35624} is caused by development code changes.
  }%
  \label{fig:intro-commits}
\end{figure}

\paragraph{Continuous Benchmarking}
aims to continuously detect performance changes in the
target system using performance measurements~\cite{Reichelt2019,EDivisive2020}.
MooBench is designed to continuously benchmark the overhead of tracing agents,
initially on Jenkins~\cite{Waller2015} using a bare metal server, and extended
to GitHub actions~\cite{Overhead2024} using GitHub-hosted VMs. The
github-action-benchmark is a plugin application native to the GitHub CI/CD,
enabling collection and visualization of data from continuous benchmarking.
Nyrki\"{o} extends github-action-benchmark, incorporating the E-Divisive
algorithm to detect performance changes, and utilizing the GitHub Issues board
to notify important detection results.

\paragraph{E-Divisive Means Algorithm in Nyrki\"{o}\label{sec:nyrkio}}

\begin{figure}[t]
  \centering
  \includegraphics[width=\linewidth,trim={0 0 0 0},clip=true]
  {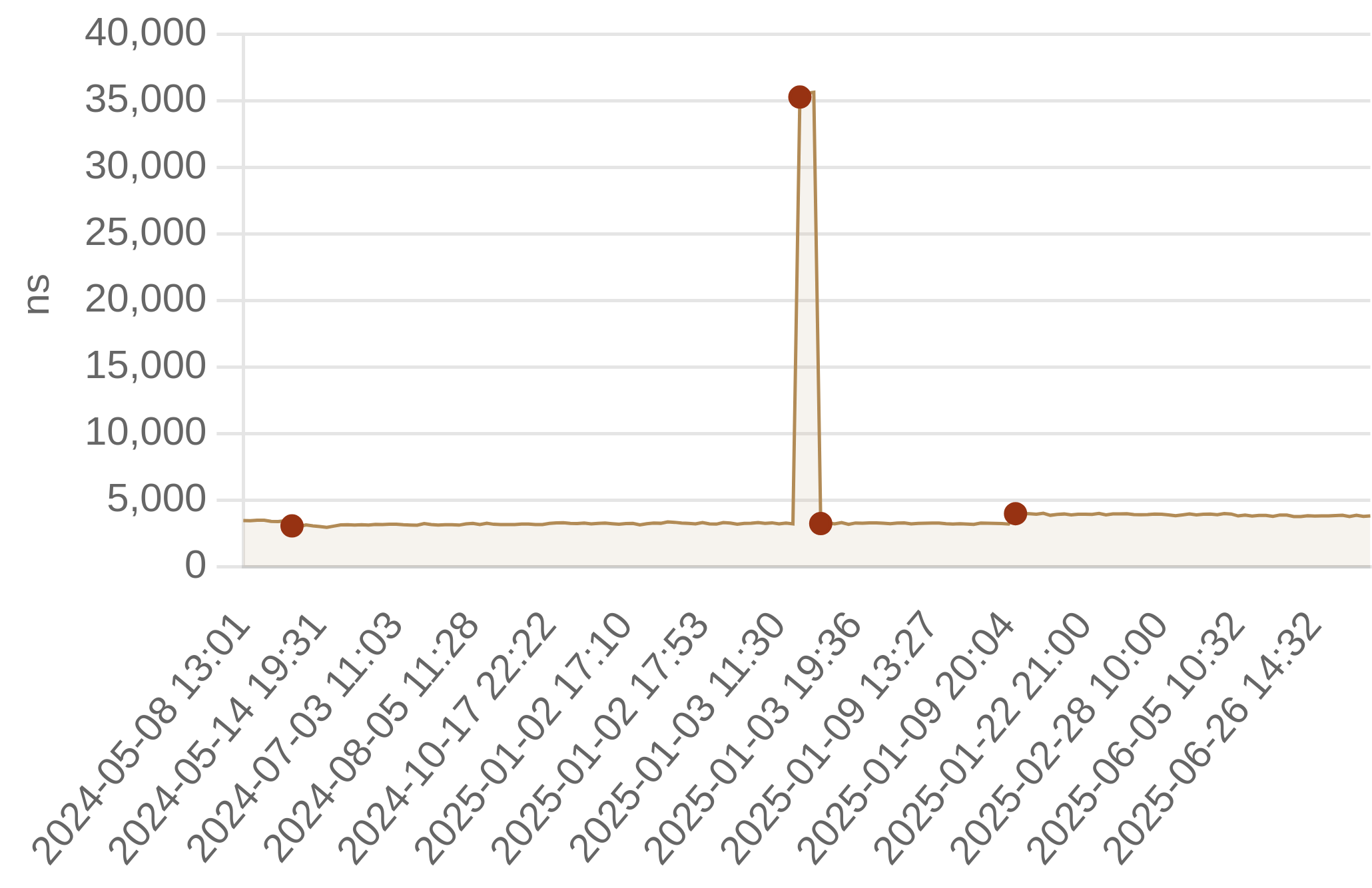}
  \caption[Commit Diagram]
  {%
    Nyrki\"{o}'s detected performance changes
  }%
  \label{fig:nyrkio-result}
  \vspace{-1.00em}
\end{figure}

Matteson and James first introduced the E-Divisive algorithm, which detects
change points in the given numerical sequence~\cite{EDivisive2014}.
We used Nyrki\"{o}'s E-Divisive implementation to validate that our initial finding on the performance change between two commits as shown in
Figure~\ref{fig:intro-commits}. In Figure~\ref{fig:nyrkio-result}, produced by Nyrki\"{o}, we see that our finding is indeed recognized as a performance change by E-Divisive Means: the diagram shows four change detections in four red dots, and the right-most dot on January 9, 2025 at 20:04 matches the performance change between {\tt cf3c08d} and {\tt 59d51e3} in Figure~\ref{fig:intro-commits}. To produce it, we configured the p-value with \num{0.001},
which finds fewer change points, decreasing false positives. Setting the change
magnitude with \SI{5}{\percent}, Nyrki\"{o} only lists change points bigger than
that.

\section{Performance Change Examination}

In this section, we present the performance change detection, our experiments for examining one change and the analysis of the experimental data.

\subsection{Previously Detected Changes}

In May 2024, MooBench started continuous benchmarking on the GitHub
CI/CD~\cite{Overhead2024}.  In addition to git-push, GitHub actions allow for
triggering a CI/CD pipeline with a scheduler to collect the data periodically.
In March 2025, we noticed a performance regression of Kieker's AspectJ agent from
the GitHub CI/CD results.
The regression appeared between two benchmark runs triggered by two git-pushes,
noted by two commit signatures, {\tt cf3c08d} and {\tt 59d51e3}. We made those
pushes to install the R package, which was pre-installed in the old {\tt
ubuntu-22.04} image, but obsoleted in the new {\tt ubuntu-24.04} image, a
decision by GitHub.%
\footnote{\url{https://github.com/actions/runner-images/issues/10636}}
Our first intuition was that the performance regression was made by the Ubuntu
version changes. 

\subsection{Experimental Setup\label{sec:exp}}

To replicate the performance change between two commit signatures {\tt cf3c08d}
and {\tt 59d51e3}, we deployed eight runners in our GitHub Actions workflow:
four GitHub-hosted runners and four self-hosted runners. All eight VMs were
configured to start either by a git-push, or a scheduler that starts them every
three hours.

{\bf GitHub-hosted runners}:
we use four standard GitHub-hosted runners. Each deploys a Linux image: two
runners use the {\tt ubuntu-22.04} image, and the other two use the {\tt
ubuntu-24.04} image. Kozlov reports that a standard GitHub-hosted runner has
4~vCPUs and \SI{16}{\gibi\byte} RAM, which is equivalent to the {\tt t4g.xlarge}
instance on AWS~\cite{Kozlov2025}.

{\bf Self-hosted runners}:
we use four self-hosted runners. Our
self-hosted runners are four virtual machines with an identical spec; each has
4~vCPUs and \SI{8}{\gibi\byte} RAM. The VMs belong to the same hosting server,
which has 2~Xeon E5-2650 CPUs and \SI{94}{\gibi\byte} RAM. Two of them use
Ubuntu 22.04 and the other two use Ubuntu 24.04.

{\bf MooBench}:
We benchmarked the Kieker AspectJ agent, where we observed the performance
change. MooBench uses its default configuration to run the Kieker AspectJ agent:
one iteration measures the duration of a Java method, which recursively calls
itself $10$~times. One loop includes two million iterations, and the entire
benchmark consists of $10$~loops, and the system rests for \SI{30}{\second} at
the end of each loop.

\subsection{Statistical Analysis}

\paragraph{Evaluation method:} For all pairs of compared data series, we
checked their normality using the Shapiro-Wilk normality test. The resulting
$w$-value close to \num{1.00} confirms the normality of the compared data. We
used the paired t-test to verify whether the difference of the two compared data
series is statistically significant or not. The resulting $p$-value $<$
\num{0.05} means the difference is significant.

\begin{figure}[t]
  \centering
  \hfill 
\begin{tikzpicture}[x=1pt,y=1pt]

\def\LEFT{0}
\def\TOP{0}
\def\ColI{\LEFT+19.70}
\def\ColII{\ColI+43}
\def\ColIII{\ColII+43}
\def\ColIV{\ColIII+58}
\def\RowHeight{8.3}
\def\RowI{\TOP+13.5}
\def\RowII{\RowI+\RowHeight}
\def\RowIII{\RowII+\RowHeight}
\def\RowIV{\RowIII+\RowHeight}
\def\RowV{\RowIV+\RowHeight}
\def\RowVI{\RowV+\RowHeight}
\def\RowVII{\RowVI+\RowHeight}
\def\RowVIII{\RowVII+\RowHeight}
\def\GuideX{26.47}

\definecolor{fillColor}{RGB}{255,255,255}
\path[use as bounding box,fill=fillColor,fill opacity=0.00] (0,0) rectangle (224.04, 82.39);

\begin{scope}
\definecolor{drawColor}{RGB}{166,206,227}
\definecolor{fillColor}{RGB}{166,206,227}

\path[draw=drawColor,line width= 0.4pt,line join=round,line cap=round,
  fill=fillColor] (\ColI, \RowI) circle (2);
\end{scope}

\begin{scope}
\definecolor{drawColor}{RGB}{31,120,180}
\definecolor{fillColor}{RGB}{31,120,180}

\path[draw=drawColor,line width= 0.4pt,line join=round,line cap=round,
  fill=fillColor] (\ColI, \RowII) circle (2);
\end{scope}

\begin{scope}
\definecolor{drawColor}{RGB}{178,223,138}
\definecolor{fillColor}{RGB}{178,223,138}

\path[draw=drawColor,line width= 0.4pt,line join=round,line cap=round,
  fill=fillColor] (\ColI, \RowIII) circle (2);
\end{scope}

\begin{scope}
\definecolor{drawColor}{RGB}{51,160,44}
\definecolor{fillColor}{RGB}{51,160,44}

\path[draw=drawColor,line width= 0.4pt,line join=round,line cap=round,
  fill=fillColor] (\ColI, \RowIV) circle (2);
\end{scope}

\begin{scope}
\definecolor{drawColor}{RGB}{251,154,153}
\definecolor{fillColor}{RGB}{251,154,153}

\path[draw=drawColor,line width= 0.4pt,line join=round,line cap=round,
  fill=fillColor] (\ColI, \RowV) circle (2);
\end{scope}

\begin{scope}
\definecolor{drawColor}{RGB}{227,26,28}
\definecolor{fillColor}{RGB}{227,26,28}

\path[draw=drawColor,line width= 0.4pt,line join=round,line cap=round,
  fill=fillColor] (\ColI, \RowVI) circle (2);
\end{scope}

\begin{scope}
\definecolor{drawColor}{RGB}{253,191,111}
\definecolor{fillColor}{RGB}{253,191,111}

\path[draw=drawColor,line width= 0.4pt,line join=round,line cap=round,
  fill=fillColor] (\ColI, \RowVII) circle (2);
\end{scope}

\begin{scope}
\definecolor{drawColor}{RGB}{255,127,0}
\definecolor{fillColor}{RGB}{255,127,0}

\path[draw=drawColor,line width= 0.4pt,line join=round,line cap=round,
  fill=fillColor] (\ColI, \RowVIII) circle (2);
\end{scope}

\begin{scope}
\path[clip] ( \GuideX,  8.25) rectangle (218.54, 80.24);
\definecolor{drawColor}{gray}{0.92}

\path[draw=drawColor,line width= 0.3pt,line join=round] ( 73.86,  8.25) --
	( 73.86, 76.89);

\path[draw=drawColor,line width= 0.3pt,line join=round] (127.64,  8.25) --
	(127.64, 76.89);

\path[draw=drawColor,line width= 0.3pt,line join=round] (181.41,  8.25) --
	(181.41, 76.89);

\path[draw=drawColor,line width= 0.6pt,line join=round] ( \GuideX, 13.27) --
	(218.54, 13.27);

\path[draw=drawColor,line width= 0.6pt,line join=round] ( \GuideX, 21.64) --
	(218.54, 21.64);

\path[draw=drawColor,line width= 0.6pt,line join=round] ( \GuideX, 30.01) --
	(218.54, 30.01);

\path[draw=drawColor,line width= 0.6pt,line join=round] ( \GuideX, 38.38) --
	(218.54, 38.38);

\path[draw=drawColor,line width= 0.6pt,line join=round] ( \GuideX, 46.75) --
	(218.54, 46.75);

\path[draw=drawColor,line width= 0.6pt,line join=round] ( \GuideX, 55.12) --
	(218.54, 55.12);

\path[draw=drawColor,line width= 0.6pt,line join=round] ( \GuideX, 63.50) --
	(218.54, 63.50);

\path[draw=drawColor,line width= 0.6pt,line join=round] ( \GuideX, 71.87) --
	(218.54, 71.87);

\path[draw=drawColor,line width= 0.6pt,line join=round] ( 46.98,  8.25) --
	( 46.98, 76.89);

\path[draw=drawColor,line width= 0.6pt,line join=round] (100.75,  8.25) --
	(100.75, 76.89);

\path[draw=drawColor,line width= 0.6pt,line join=round] (154.52,  8.25) --
	(154.52, 76.89);

\path[draw=drawColor,line width= 0.6pt,line join=round] (208.29,  8.25) --
	(208.29, 76.89);
\definecolor{fillColor}{RGB}{166,206,227}

\path[fill=fillColor] (183.96, 13.27) circle (  0.43);
\definecolor{drawColor}{RGB}{166,206,227}

\path[draw=drawColor,line width= 0.6pt,line join=round] (160.95, 13.27) -- (179.93, 13.27);

\path[draw=drawColor,line width= 0.6pt,line join=round] (147.95, 13.27) -- (130.85, 13.27);

\path[draw=drawColor,line width= 0.6pt,fill=fillColor] (160.95, 10.13) --
	(155.02, 10.13) --
	(152.39, 11.70) --
	(149.76, 10.13) --
	(147.95, 10.13) --
	(147.95, 16.41) --
	(149.76, 16.41) --
	(152.39, 14.84) --
	(155.02, 16.41) --
	(160.95, 16.41) --
	(160.95, 10.13) --
	cycle;

\path[draw=drawColor,line width= 1.1pt] (152.39, 11.70) -- (152.39, 14.84);
\definecolor{drawColor}{RGB}{31,120,180}

\path[draw=drawColor,line width= 0.6pt,line join=round] (159.60, 21.64) -- (176.40, 21.64);

\path[draw=drawColor,line width= 0.6pt,line join=round] (145.57, 21.64) -- (126.89, 21.64);
\definecolor{fillColor}{RGB}{31,120,180}

\path[draw=drawColor,line width= 0.6pt,fill=fillColor] (159.60, 18.50) --
	(153.44, 18.50) --
	(150.58, 20.07) --
	(147.72, 18.50) --
	(145.57, 18.50) --
	(145.57, 24.78) --
	(147.72, 24.78) --
	(150.58, 23.21) --
	(153.44, 24.78) --
	(159.60, 24.78) --
	(159.60, 18.50) --
	cycle;

\path[draw=drawColor,line width= 1.1pt] (150.58, 20.07) -- (150.58, 23.21);
\definecolor{fillColor}{RGB}{178,223,138}

\path[fill=fillColor] (161.93, 30.01) circle (  0.43);
\definecolor{drawColor}{RGB}{178,223,138}

\path[draw=drawColor,line width= 0.6pt,line join=round] (191.60, 30.01) -- (205.16, 30.01);

\path[draw=drawColor,line width= 0.6pt,line join=round] (180.06, 30.01) -- (163.23, 30.01);

\path[draw=drawColor,line width= 0.6pt,fill=fillColor] (191.60, 26.87) --
	(189.65, 26.87) --
	(187.31, 28.44) --
	(184.98, 26.87) --
	(180.06, 26.87) --
	(180.06, 33.15) --
	(184.98, 33.15) --
	(187.31, 31.58) --
	(189.65, 33.15) --
	(191.60, 33.15) --
	(191.60, 26.87) --
	cycle;

\path[draw=drawColor,line width= 1.1pt] (187.31, 28.44) -- (187.31, 31.58);
\definecolor{drawColor}{RGB}{51,160,44}

\path[draw=drawColor,line width= 0.6pt,line join=round] (195.44, 38.38) -- (207.66, 38.38);

\path[draw=drawColor,line width= 0.6pt,line join=round] (178.57, 38.38) -- (163.30, 38.38);
\definecolor{fillColor}{RGB}{51,160,44}

\path[draw=drawColor,line width= 0.6pt,fill=fillColor] (195.44, 35.24) --
	(190.94, 35.24) --
	(187.53, 36.81) --
	(184.11, 35.24) --
	(178.57, 35.24) --
	(178.57, 41.52) --
	(184.11, 41.52) --
	(187.53, 39.95) --
	(190.94, 41.52) --
	(195.44, 41.52) --
	(195.44, 35.24) --
	cycle;

\path[draw=drawColor,line width= 1.1pt] (187.53, 36.81) -- (187.53, 39.95);
\definecolor{drawColor}{RGB}{251,154,153}

\path[draw=drawColor,line width= 0.6pt,line join=round] ( 44.79, 46.75) -- ( 46.52, 46.75);

\path[draw=drawColor,line width= 0.6pt,line join=round] ( 42.97, 46.75) -- ( 41.58, 46.75);
\definecolor{fillColor}{RGB}{251,154,153}

\path[draw=drawColor,line width= 0.6pt,fill=fillColor] ( 44.79, 43.62) --
	( 44.33, 43.62) --
	( 43.91, 45.18) --
	( 43.49, 43.62) --
	( 42.97, 43.62) --
	( 42.97, 49.89) --
	( 43.49, 49.89) --
	( 43.91, 48.32) --
	( 44.33, 49.89) --
	( 44.79, 49.89) --
	( 44.79, 43.62) --
	cycle;

\path[draw=drawColor,line width= 1.1pt] ( 43.91, 45.18) -- ( 43.91, 48.32);
\definecolor{drawColor}{RGB}{227,26,28}

\path[draw=drawColor,line width= 0.6pt,line join=round] ( 44.70, 55.12) -- ( 47.01, 55.12);

\path[draw=drawColor,line width= 0.6pt,line join=round] ( 42.65, 55.12) -- ( 40.65, 55.12);
\definecolor{fillColor}{RGB}{227,26,28}

\path[draw=drawColor,line width= 0.6pt,fill=fillColor] ( 44.70, 51.99) --
	( 44.24, 51.99) --
	( 43.82, 53.56) --
	( 43.41, 51.99) --
	( 42.65, 51.99) --
	( 42.65, 58.26) --
	( 43.41, 58.26) --
	( 43.82, 56.69) --
	( 44.24, 58.26) --
	( 44.70, 58.26) --
	( 44.70, 51.99) --
	cycle;

\path[draw=drawColor,line width= 1.1pt] ( 43.82, 53.56) -- ( 43.82, 56.69);
\definecolor{drawColor}{RGB}{253,191,111}

\path[draw=drawColor,line width= 0.6pt,line join=round] ( 45.50, 63.50) -- ( 48.16, 63.50);

\path[draw=drawColor,line width= 0.6pt,line join=round] ( 43.42, 63.50) -- ( 41.35, 63.50);
\definecolor{fillColor}{RGB}{253,191,111}

\path[draw=drawColor,line width= 0.6pt,fill=fillColor] ( 45.50, 60.36) --
	( 45.18, 60.36) --
	( 44.80, 61.93) --
	( 44.43, 60.36) --
	( 43.42, 60.36) --
	( 43.42, 66.63) --
	( 44.43, 66.63) --
	( 44.80, 65.06) --
	( 45.18, 66.63) --
	( 45.50, 66.63) --
	( 45.50, 60.36) --
	cycle;

\path[draw=drawColor,line width= 1.1pt] ( 44.80, 61.93) -- ( 44.80, 65.06);
\definecolor{fillColor}{RGB}{255,127,0}

\path[fill=fillColor] ( 40.82, 71.87) circle (  0.43);

\path[fill=fillColor] ( 39.44, 71.87) circle (  0.43);
\definecolor{drawColor}{RGB}{255,127,0}

\path[draw=drawColor,line width= 0.6pt,line join=round] ( 45.56, 71.87) -- ( 47.77, 71.87);

\path[draw=drawColor,line width= 0.6pt,line join=round] ( 43.81, 71.87) -- ( 41.98, 71.87);

\path[draw=drawColor,line width= 0.6pt,fill=fillColor] ( 45.56, 68.73) --
	( 45.17, 68.73) --
	( 44.82, 70.30) --
	( 44.46, 68.73) --
	( 43.81, 68.73) --
	( 43.81, 75.00) --
	( 44.46, 75.00) --
	( 44.82, 73.43) --
	( 45.17, 75.00) --
	( 45.56, 75.00) --
	( 45.56, 68.73) --
	cycle;

\path[draw=drawColor,line width= 1.1pt] ( 44.82, 70.30) -- ( 44.82, 73.43);
\end{scope}
\begin{scope}
\path[clip] (  0.00,  0.00) rectangle (224.04, 82.39);
\definecolor{drawColor}{RGB}{0,0,0}

\node[text=drawColor,rotate= 90.00,anchor=base,inner sep=0pt, outer sep=0pt, scale=  0.90] at (  6.20, 42.57) {Runtime};

\node[text=drawColor,rotate= 90.00,anchor=base,inner sep=0pt, outer sep=0pt, scale=  0.90] at ( 14.32, 42.57) {Environments};
\end{scope}
\end{tikzpicture}
\begin{tikzpicture}[x=1pt,y=1pt]

\def\LEFT{0}
\def\TOP{0}
\def\ColI{\LEFT+36}
\def\ColII{\ColI+43}
\def\ColIII{\ColII+43}
\def\ColIV{\ColIII+58}
\def\RowI{\TOP+19.95}
\def\RowII{\TOP+9.95}

\begin{scope}
\definecolor{drawColor}{RGB}{166,206,227}
\definecolor{fillColor}{RGB}{166,206,227}

\path[draw=drawColor,line width= 0.4pt,line join=round,line cap=round,
  fill=fillColor] (\ColI, \RowI) circle (2);
\end{scope}

\begin{scope}
\definecolor{drawColor}{RGB}{31,120,180}
\definecolor{fillColor}{RGB}{31,120,180}

\path[draw=drawColor,line width= 0.4pt,line join=round,line cap=round,
  fill=fillColor] (\ColI, \RowII) circle (2);
\end{scope}

\begin{scope}
\definecolor{drawColor}{RGB}{178,223,138}
\definecolor{fillColor}{RGB}{178,223,138}

\path[draw=drawColor,line width= 0.4pt,line join=round,line cap=round,
  fill=fillColor] (\ColII, \RowI) circle (2);
\end{scope}

\begin{scope}
\definecolor{drawColor}{RGB}{51,160,44}
\definecolor{fillColor}{RGB}{51,160,44}

\path[draw=drawColor,line width= 0.4pt,line join=round,line cap=round,
  fill=fillColor] (\ColII, \RowII) circle (2);
\end{scope}

\begin{scope}
\definecolor{drawColor}{RGB}{251,154,153}
\definecolor{fillColor}{RGB}{251,154,153}

\path[draw=drawColor,line width= 0.4pt,line join=round,line cap=round,
  fill=fillColor] (\ColIII, \RowI) circle (2);
\end{scope}

\begin{scope}
\definecolor{drawColor}{RGB}{227,26,28}
\definecolor{fillColor}{RGB}{227,26,28}

\path[draw=drawColor,line width= 0.4pt,line join=round,line cap=round,
  fill=fillColor] (\ColIII, \RowII) circle (2);
\end{scope}

\begin{scope}
\definecolor{drawColor}{RGB}{253,191,111}
\definecolor{fillColor}{RGB}{253,191,111}

\path[draw=drawColor,line width= 0.4pt,line join=round,line cap=round,
  fill=fillColor] (\ColIV, \RowI) circle (2);
\end{scope}

\begin{scope}
\definecolor{drawColor}{RGB}{255,127,0}
\definecolor{fillColor}{RGB}{255,127,0}

\path[draw=drawColor,line width= 0.4pt,line join=round,line cap=round,
  fill=fillColor] (\ColIV, \RowII) circle (2);
\end{scope}

\begin{scope}

\definecolor{drawColor}{RGB}{0,0,0}

\node[text=drawColor,anchor=base west,inner sep=0pt, outer sep=0pt, scale= 0.80]
  at (\LEFT, \RowI-2.03) {{\tt cf3c08d}:}; \end{scope} \begin{scope} \definecolor{drawColor}{RGB}{0,0,0}
\node[text=drawColor,anchor=base west,inner sep=0pt, outer sep=0pt, scale= 0.80]
  at (\LEFT, \RowII-2.03) {{\tt 59d51e3}:}; \end{scope} \begin{scope} \definecolor{drawColor}{RGB}{0,0,0}

\node[text=drawColor,anchor=base west,inner sep=0pt, outer sep=0pt, scale= 0.80]
  at (\ColI+3, \RowI-3.03) {22.04/self}; \end{scope} \begin{scope} \definecolor{drawColor}{RGB}{0,0,0}
\node[text=drawColor,anchor=base west,inner sep=0pt, outer sep=0pt, scale= 0.80]
  at (\ColI+3, \RowII-3.03) {22.04/self}; \end{scope} \begin{scope} \definecolor{drawColor}{RGB}{0,0,0}

\node[text=drawColor,anchor=base west,inner sep=0pt, outer sep=0pt, scale= 0.80]
  at (\ColII+3, \RowI-3.03) {24.04/self}; \end{scope} \begin{scope} \definecolor{drawColor}{RGB}{0,0,0}
\node[text=drawColor,anchor=base west,inner sep=0pt, outer sep=0pt, scale= 0.80]
  at (\ColII+3, \RowII-3.03) {24.04/self}; \end{scope} \begin{scope} \definecolor{drawColor}{RGB}{0,0,0}

\node[text=drawColor,anchor=base west,inner sep=0pt, outer sep=0pt, scale= 0.80]
  at (\ColIII+3, \RowI-3.03) {22.04/GitHub}; \end{scope} \begin{scope} \definecolor{drawColor}{RGB}{0,0,0}
\node[text=drawColor,anchor=base west,inner sep=0pt, outer sep=0pt, scale= 0.80]
  at (\ColIII+3, \RowII-3.03) {22.04/GitHub}; \end{scope} \begin{scope} \definecolor{drawColor}{RGB}{0,0,0}

\node[text=drawColor,anchor=base west,inner sep=0pt, outer sep=0pt, scale= 0.80]
  at (\ColIV+3, \RowI-3.03) {24.04/GitHub}; \end{scope} \begin{scope} \definecolor{drawColor}{RGB}{0,0,0}
\node[text=drawColor,anchor=base west,inner sep=0pt, outer sep=0pt, scale= 0.80]
  at (\ColIV+3, \RowII-3.03) {24.04/GitHub}; \end{scope}
\end{tikzpicture}
  \vspace{-1.00em}
  \caption[Validation Diagram]
  {%
    Measurements at commits {\tt
    cf3c08d}~(\fourlines{Paired0}{Paired2}{Paired4}{Paired6}) and {\tt
    59d51e3}~(\fourlines{Paired1}{Paired3}{Paired5}{Paired7}). The boxplot shows
    the distribution of execution time measurements per runtime environment.
  }%
  \label{fig:validation}
  \vspace{-1.00em}
\end{figure}
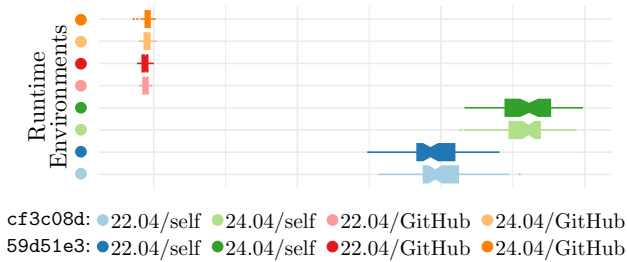

\paragraph{Validation:} We elaborated the results in
Figure~\ref{fig:validation}. First, we validated that the software change was
not the cause of the performance change.
The $\left(w,p\right)$ between the two execution time series by commits {\tt
cf3c08d} and {\tt 59d51e3} are
$\left(\float{0.96246}, \float{0.07228757}\right)$ (22.04/self),
$\left(\float{0.98334}, \float{0.6479095} \right)$ (24.04/self),
$\left(\float{0.97852}, \float{0.6480887} \right)$ (22.04/GitHub), and
$\left(\float{0.98486}, \float{0.4884971} \right)$ (24.04/GitHub).
Using our method, the compared data series are normally
distributed, and the difference of comparisons are not significant.
Second, we validated our assumption both on GitHub-hosted runners and
self-hosted runners by comparing two
execution time series by two Ubuntu versions \num{22.04} and \num{24.04}:
$\left(\float{0.98828}, \num{0.001966284}\right)$ (GitHub-hosted) and
$\left(\float{0.98859}, \num{1.368409e-50}\right)$ (self-hosted).
In the future, we will investigate Ubuntu software changes and evaluate the
differences, and also validate whether this is related to different kernel
versions.

\section{Related Work}
The E-Divisive means algorithm was introduced in~\cite{EDivisive2014}.
The efforts to integrate E-Divisive means algorithms to CI/CD has
followed~\cite{EDivisive2020, EDivisive2023, EDivisive2025}.
Regression testing in continuous integration is investigated
in~\cite{Reichelt2019}. They utilize unit testing for regression testing, which
compares two different code versions.
Kozlov~\cite{Kozlov2025} did a comprehensive analysis on GitHub-hosted VMs
and compared them to the equivalent AWS EC2 instances, and self-hosted VMs.
Nyrki\"{o}'s GitHub action tool\footnote{\url{https://github.com/nyrkio/github-action-benchmark}} is based on the github-action-benchmark tool.
Extending the MooBench CI/CD to GitHub and the use of github-action-benchmark
discussed in~\cite{Reichelt2024}.

\section{Conclusion and Future Work}
In this work, we discussed how to continuously detect performance changes
for MooBench. Using the Nyrki\"{o} change detection, we identified a significant performance change.
To show the performance change does not only occur on GitHub-hosted VMs, we replicated the performance
change on self-hosted VMs in our local servers too.
In the future, we will further investigate on the performance change, and use the updated MooBench GitHub workflow to examine
regressions caused by source code changes of the tracing agents.

\paragraph{Acknowledgment}
This research is funded by the Deutsche Forschungsgemeinschaft (DFG -- German
Research Foundation), grant no.~528713834.

\printbibliography

\end{document}